\documentclass[aps,pre,showpacs,amsmath,amsfonts,amssymb,superscriptaddress]{revtex4} 
\usepackage[dvips]{graphicx}
\usepackage{bbm}

\newcommand{\beq}{\begin{equation}} 
\newcommand{\eeq}{\end{equation}}
\newcommand{\bea}{\begin{eqnarray}} 
\newcommand{\eea}{\end{eqnarray}}

\input epsf 
 
\begin{document} 
 
\title{Geometry of the energy landscape of the self-gravitating ring} 

\author{Bernardo Monechi} 
\email{mone.berna@gmail.com} 
\affiliation{Dipartimento di Fisica, ``Sapienza'' Universit\`a di Roma, piazzale A.\ Moro 2, I-00185 Roma, Italy}  
\author{Lapo Casetti} 
\email{lapo.casetti@unifi.it} 
\affiliation{Dipartimento di Fisica e Astronomia and Centro per lo Studio delle Dinamiche Complesse (CSDC), Universit\`a di Firenze, via G.~Sansone 1, I-50019 Sesto Fiorentino (FI), Italy}  
\affiliation{Istituto Nazionale di Fisica Nucleare (INFN), Sezione di Firenze, via G.~Sansone 1, I-50019 Sesto Fiorentino (FI), Italy}  

\date{\today} 
 
\begin{abstract} 
We study the global geometry of the energy landscape of a simple model of a self-gravitating system, the self-gravitating ring (SGR). This is done by endowing the configuration space with a metric such that the dynamical trajectories are identified with geodesics. The average curvature and curvature fluctuations of the energy landscape are computed by means of Monte Carlo simulations and, when possible, of a mean-field method, showing that these global geometric quantities provide a clear geometric characterization of the collapse phase transition occurring in the SGR as the transition from a flat landscape at high energies to a landscape with mainly positive but fluctuating curvature in the collapsed phase. Moreover, curvature fluctuations show a maximum in correspondence with the energy of a possible further transition, occurring at lower energies than the collapse one, whose existence had been previously conjectured on the basis of a local analysis of the energy landscape and whose effect on the usual thermodynamic quantities, if any, is extremely weak. We also estimate the largest Lyapunov exponent $\lambda$ of the SGR using the geometric observables. The geometric estimate always gives the correct order of magnitude of $\lambda$ and is also quantitatively correct at small energy densities and, in the limit $N\to\infty$, in the whole homogeneous phase. 
\end{abstract} 
 
 
 
\maketitle 

\section{Introduction} 
\label{sec_intro}

Given a system of classical interacting particles described by the Hamiltonian 
\beq
{\cal H}=\sum _{i=1}^{N}\frac{p_i^2}{2m} + V\left(q_1,\ldots,q_N \right)\,,
\label{H}
\eeq
all the information on the dynamics as well as on the equilibrium collective properties of the system is encoded in the potential energy $V$. Indeed, the forces entering the equations of motion are given by the gradient of $V$, and the equilibrium statistical measures in both the canonical ad the microcanonical ensembles can be defined in terms of $V$, since the kinetic energy contribution can be integrated out explicitly. 

To extract information from the function $V$ one may use a class of methods commonly referred to as ``energy landscape'' methods \cite{Wales:book}. The value of the function $V(q)$ at a given configuration $q = \left(q_1,\ldots,q_N \right) \in M$, where $M$ is the $N$-dimensional configuration space, is the height of the landscape, so that configurations with low energy are valleys (and their bottoms are minima of $V$), maxima of $V$ correspond to peaks, an saddle points of $V$ are mountain passes\footnote{More precisely saddles are classified according to their \textit{index}, i.e., by the number of unstable directions: what one would call ``proper'' mountain passes are index-one saddles.}. Hence the topography of the energy landscape is determined by the stationary points of $V$, i.e.\ the configurations $q^s$ such that $\nabla V(q^s) = 0$. Starting from the pioneering work by Stillinger and Weber \cite{StillingerWeber:science1984}, the energy landscape approach has been fruitfully applied to study glassy systems (see e.g.\ \cite{Sciortino:jstat2005} and references therein) and biologically motivated problems such as protein folding (see e.g.\ \cite{OnuchicLutheyWolynes:arpc1997} and references therein). The basic idea behind the energy landscape\footnote{A ``free energy landscape'' can be also defined by projecting the $N$-dimensional configuration space onto a small set of collective variables \protect\cite{Wales:book}.} methods is very simple, yet powerful: if a system has a rugged, complex energy landscape, with many minima and valleys separated by barriers of different height, its dynamics will experience a variety of time scales, with oscillations in the valleys and jumps from one valley to another. One of the drawbacks of this approach is that one should in principle find {\em all} the stationary points of $V$ to obtain a complete characterization of the landscape, or at least all the minima and all the saddles connecting them: for many-particle systems and generic potential energies this is an impossible task. Also assuming that a partial sampling of the landscape is sufficient, a huge computational effort is required in order to find a non-negligible fraction of minima and saddles. It is then natural to ask whether a characterization of the energy landscape can be obtained in terms of {\em global} quantities, averaged on the landscape itself (see \cite{prl2006_2,pre2008,book2009} for a related discussion in the context of toy models of proteinlike polymers).

The energy landscape approach, with its emphasis on the topography of the basins around minima, has a topological flavour\footnote{This can be made explicit by noting that at any stationary point $q^s$ of the potential energy $V$ the topology of the submanifolds of configuration space $M_v = \left\{q \in M | V(q) \le v \right\}$ changes, in a way that is completely determined by the stationary point $q^s$ itself, according to Morse theory \protect\cite{Milnor:book}. Hence whenever the energy crosses a stationary value $v^s = V(q^s)$ the topology of the part of configuration space whose energy is smaller than $v^s$ changes. It has been conjectured that some of these topological changes may be related to equilibrium phase transitions (see e.g.\ \protect\cite{physrep2000,chaos2005,Pettini:book,Kastner:rmp2008} and references therein, and also the discussion in Sec.\ \protect\ref{sec_collective}).}. However, the potential energy $V$ can also induce a geometric structure on the configuration space: a metric function $g$ on $M$ can be defined in terms of the potential energy $V$ such that the configuration space becomes a Riemannian, or pseudo-Riemannian, manifold. We can thus speak of the geometry of the energy landscape, and as we shall see in the following the averages of geometric quantities can provide the global characterization of the landscape we are looking for. The intuitive reason why geometric information on the landscape, and especially {\em curvature}, could be a relevant one is that the dynamics on a landscape would be heavily affected by the local curvature: minima of the energy landscape are associated to positive curvatures and stable dynamics, while saddles involve negative curvatures, at least along some direction, thus implying some instability. One can reasonably expect that the arrangement and detailed properties of minima and saddles might reflect in some global feature of the distribution of curvatures of the landscape, when averaged along a typical trajectory.

There are many possible ways to define a metric structure on the energy landscape. The most immediate choice would probably be that of considering as our manifold $M$ not the configuration space but the $N$-dimensional surface $z = V(q_1,\ldots,q_N)$ itself, i.e., the graph of the potential energy $V$ as a function of the $N$ coordinates $q_1,\ldots,q_N$ of the configuration space, and to define $g$ as the metric induced on that surface by its immersion in $\mathbb{R}^{N+1}$. Although perfectly reasonable, this choice has two major drawbacks. First, the explicit expressions for the geometric quantities in terms of $V$ and its derivatives are rather complicated. Second, and most relevant to what we are going to discuss in the following, with that choice of the metric the link between the properties of the dynamics and the geometry is not very precise, i.e., one cannot prove that the geometry completely determines the dynamics and its stability. We shall thus choose a metric $g$, referred to as the Eisenhart metric \cite{Eisenhart:annmath1929}, such that the dynamical trajectories are the geodesics of the configuration space endowed with such metric and that the stability of the dynamics is completely determined by the curvature of such metric. In addition, the explicit expressions for the curvature of the Eisenhart metric in terms of derivatives of $V$ are particularly simple (see Sec.\ \ref{sec_geom}). As reviewed in \cite{physrep2000,chaos2005,Pettini:book}, with such metric one can define a suitable curvature observable whose average and fluctuation allow to characterize the global properties of the energy landscape as well as to estimate the degree of instability of the dynamics, quantified by the largest Lyapunov exponent.

The aim of this paper is to apply these geometric tools to characterize the energy landscape and the dynamics of a simple model of a self-gravitating system, the self-gravitating ring (SGR), introduced in \cite{SotaEtAl:pre2001}. The microcanonical statistical mechanics of this model has been studied by a mean-field technique in \cite{TatekawaEtAl:pre2005}, showing that the system undergoes a phase transition between a collapsed and a uniform phase. The latter phase transition has been related to some particular stationary points of the energy landscape in \cite{prerap2009}, where also some indication of the possibility of another phase transition, occurring at lower energies, was presented. In the following, we shall compute the curvature of the energy landscape defined via the Eisenhart metric and show that its average and fluctuations clearly mark the collapse phase transitions; moreover, the fluctuations of the curvature also display a peak at the energy where the second phase transition conjectured in  \cite{prerap2009} should occur. We shall also estimate the largest Lyapunov exponent in terms of the average and fluctuations of the curvature and compare such estimates with direct numerical calculations. The geometric quantities provide a reasonable estimate of the Lyapunov exponents for any energy, and an accurate one for large and small energies; at intermediate energies the geometric estimate, though still giving the correct order of magnitude, is worse and we shall discuss the origin of this result.

The paper is organized as follows: in Sec.\ \ref{sec_geom} we shall briefly recall the relation between geometry and dynamics, in order to define the geometric quantities that will be studied to characterize the global properties of the landscape and that will be used to estimate the largest Lyapunov exponent. In Sec.\ \ref{sec_model} we shall describe the model studied, i.e., the self-gravitating ring (SGR), and its equilibrium statistical properties; Sec.\ \ref{sec_results} will be devoted to presenting and discussing our results on the geometric quantities as well as on the Lyapunov exponents, and we shall end the paper with some concluding remarks in Sec.\ \ref{sec_conclusions}. 

\section{Geometry and dynamics}
\label{sec_geom}

That the dynamical trajectories of a classical dynamical system whose Hamiltonian is given by Eq.\ (\ref{H}) can be viewed as geodesics of the configuration space endowed with a suitable Riemannian metric is a classic result in analytical mechanics (see e.g.\ \cite{Arnold:book}). The first to conjecture that this correspondence may be relevant to the physics of many-body systems---and especially to the foundations of statistical mechanics---was probably N.\ S.\ Krylov, who in his doctoral thesis of 1942 (reprinted in \cite{Krylov:book}) suggested that the dynamical instability of the trajectories (what we nowadays call Hamiltonian chaos) may be due to the {\em negative} curvatures of the configuration manifold. Krylov's suggestion was based on earlier mathematical results \cite{Hadamard:jmpa1898,Hedlund:bams1939}; since then, geodesics flows on negatively curved manifolds have become an important chapter of ergodic theory \cite{ergodictheory:book}. However, in many dynamical systems of physical interest curvatures do not have a definite sign or even are only positive, yet the dynamics appears to be chaotic. In these cases a subtler mechanism may be at work: the oscillations of the curvature along a dynamical trajectories may induce chaos in close analogy to parametric instability, as first suggested by Pettini \cite{Pettini:pre1993}. Starting from this observation the existence of different chaotic regimes in many Hamiltonian systems was related to the curvature of the configuration space \cite{Pettini:pre1993,pre1993} and an analytical estimate of the largest Lyapunov exponent, that is the commonly used measure of the strength of chaos in a dynamical system, was derived \cite{prl1995,pre1996}. 

In the following we shall very briefly recall the main definitions and results necessary to understand the application of the geometric approach to the SGR model: in Sec.\ \ref{sec_eisenhart} we shall describe the use of the Eisenhart metric and the derivation of an effective stability equation allowing to estimate the Lyapunov exponent, while in Sec.\ \ref{sec_collective} we shall discuss how such an equation describes an effective geometry of the energy landscape. Full details and references to related work on different models can be found in the reviews \cite{physrep2000,chaos2005,Pettini:book}.

\subsection{Eisenhart metric and curvature}
\label{sec_eisenhart}

As already argued above and as it will become clear in the following, an important geometric quantity to characterize the global properties of the energy landscape is {\em curvature}. The  definition of the
curvature of a manifold $M$ depends on the choice of a metric $g$
\cite{Frankel:book,DoCarmo:book}: once the couple $(M,g)$ is given, a covariant
derivative $\nabla$ and a curvature tensor $R(e_i,e_j)$ can be defined; the latter
measures the noncommutativity of the covariant derivatives in the coordinate
directions $e_i$ and $e_j$. A scalar measure of the curvature at any given
point $q \in M$  is the  the sectional curvature 
\begin{equation}
K(e_i,e_j) = \langle R(e_i,e_j) e_j,e_i\rangle~,
\label{sec_curv}
\end{equation}
where $\langle \cdot,\cdot \rangle$ stands for the
scalar product. At any point of an $N$-dimensional manifold there are $N(N-1)$
sectional curvatures, whose knowledge determines the full curvature tensor at
that point. One can however define some  simpler
curvatures (paying the price of losing some information): 
the Ricci curvature $K_R(e_i)$ is the sum of the $K$'s over the $N-1$
directions orthogonal to $e_i$, 
\begin{equation}
K_R(e_i) = \sum_{j = 1}^N
K(e_i,e_j)~, \label{riccicurv}
\end{equation}
and summing also on the $N$ directions $e_i$ one gets the scalar
curvature 
\begin{equation}
{\cal R} = \sum_{i = 1}^N K_R(e_i) =
\sum_{i,j = 1}^N K(e_i,e_j)~;
\end{equation}  
then, $\frac{K_R}{N-1}$ and $\frac{{\cal R}}{N(N-1)}$ can be considered as
average curvatures at a given point. 

As argued in previous works (see e.g.\ \cite{physrep2000} and references therein), the Eisenhart metric \cite{Eisenhart:annmath1929} turns out to be a particularly good choice. Given a system with Hamiltonian (\ref{H}) and setting $m=1$ without loss of generality, this (pseudo-Riemannian) metric is defined on a configuration space with two extra dimensions, $M\times \mathbb{R}^2$, with local coordinates $(q_{0},q_{1},...,q_{N},q_{N+1})$, and its arc-length is
\begin{equation}
ds^{2}=\delta_{i,j}dq^{i}dq^{j}-2V(q)(dq^{0})^{2}+2dq^{0}dq^{N+1} \ . \label{11_mio}
\end{equation} 
The metric tensor will be referred to as $g_E$ and its components are 
\begin{equation}
g_E = \left( 
\begin{array}{ccccc} 
-2 V(q)& 0       & \cdots        & 0     & 1     \\ 
0       & 1& \cdots        & 0& 0     \\ 
\vdots  & \vdots& \ddots        & \vdots& \vdots\\ 
0       & 0& \cdots        & 1& 0     \\ 
1       & 0     & \cdots        & 0     & 0     \\ 
\end{array} \right) 
\label{g_E}
\end{equation}
as can be derived from Eq.\ (\ref{11_mio}).

The geodesics, i.e.\ the ``straight lines'' on the manifold, are defined by the equation
\beq
\nabla_v v = 0\,,
\label{eq_geo}
\eeq
where $v$ is the velocity vector along the geodesics itself. This means that geodesics are the curves whose (covariant) acceleration vanishes. Writing Eq.\ (\ref{eq_geo}) in local coordinates with the metric $g_E$ one finds that the geodesics coincide with the dynamical  trajectories of Hamiltonian (\ref{H}), provided one restricts to the subset of geodesics such that the parametrization of their arc-length is affine,
\beq
ds^2 = C^2 dt^2\,,
\label{affine}
\eeq
where we can set $C^2=1$ without loss of generality. The nonvanishing components of the curvature tensor are
\begin{equation}
R_{0i0j}=\frac{\partial^2 V}{\partial q_{i}\partial q_{j}}\,, 
\label{20_mio}
\end{equation} 
where $i,j=1,\ldots,N$. It can then be shown that the Ricci curvature (\ref{riccicurv}) in the direction of motion, i.e., in the direction of the velocity vector $v$ of the geodesic, is given by
\begin{equation}
K_{R}(v)=\triangle V \label{22_mio}~,
\end{equation} 
where $\triangle V$ is the Laplacian of the potential $V$, and that the scalar curvature ${\cal R}$ identically vanishes. We note that $K_{R}(v)$ is nothing but a scalar measure of the average curvature ``felt'' by the system during its evolution; we will refer to it simply as $K_R$ dropping the dependence on the direction. Another feature of $K_{R}$ is its very simple analytical expression which simplifies both analytical calculation and numerical estimates. It is also worth noticing that expression (\ref{22_mio}) is a very natural and intuitive measure of the curvature of the energy landscape, as it can be seen as a naive generalization of the curvature $f''(x)$ of the graph of a one-variable function to the graph of the $N$-dimensional function  $V(q_{1},...,q_{N})$: the Laplacian of the function. However, the previous discussion shows that it is much more than a naive measure of curvature and that it contains information on the local neighborhood of the dynamical trajectories, i.e., on their stability. As we shall see in the following, this information is sufficient to extract an estimate of the largest Lyapunov exponent of the system.

\subsubsection{Geometric estimate of the Lyapunov exponent}
\label{sec_geomlyap}

The stability of the geodesics (\ref{eq_geo}) is completely determined by the curvature of the manifold: given a geodesic whose velocity is $v$, a vector field $J$, referred to as the Jacobi field, can be defined such that $\langle J,v \rangle = 0$ and it measures the distance between nearby geodesics. The Jacobi field obeys the Jacobi equation  \cite{DoCarmo:book}
\beq
\nabla^2_v J = R(v,J)v\,.
\label{JLC}
\eeq
Written in local coordinates, Eq.\ (\ref{JLC}) becomes (summation over repeated indices is understood)
\beq
\frac{D^2 J^i}{ds^2} + R^i_{~jkl}\frac{dq^j}{ds} J^k \frac{dq^l}{ds}
\eeq
and in the particular case of the Eisenhart metric we get \cite{physrep2000}
\beq
\frac{d^2 J^i}{dt^2} + \frac{\partial^2 V}{\partial q^i \partial q^k}J^k\,,
\label{tang_dyn}
\eeq
which is commonly referred to as the tangent dynamics equation: the largest Lyapunov exponent is the mean exponential growth rate of the solutions of the latter equation. Hence using the Eisenhart metric the growth rate of the Jacobi field is directly related to the Lyapunov exponent of the system.
 
In general Eq.\ (\ref{JLC}) is very complicated. There are anyhow two special cases where it becomes simple. The first case is when the sectional curvature (\ref{sec_curv}) of the manifold is a constant; such manifolds are referred to as {\em isotropic} manifolds\footnote{These manifolds are called isotropic since the sectional curvature $K(v,w)$ at a given point $p \in M$ does not depend on the point $p$ if and only of it does not depend on the choice of the directions $v$ and $w$ \protect\cite{DoCarmo:book}.}. Choosing for simplicity a geodesic reference frame, that is a frame parallelly transported along the geodesic, the Jacobi equation can be written
\beq
\frac{d^2 J^i}{ds^2} + K J^i = 0\,,
\label{JLC_iso}
\eeq 
where $J^i$ is any of the components of $J$ in the chosen frame, $s$ is the arclength and $K$ is the constant sectional curvature. The character of the solutions depends on the sign of $K$. If $K > 0$, $J$ remains bounded and oscillates with frequency $\sqrt{K}$. If $K = 0$, $J$ grows linearly. Hence on a positive (or zero) constant curvature manifold all the geodesics are stable (or marginally stable). On the contrary, on a manifold with constant negative curvature all the geodesics are unstable, since the solutions of Eq.\ (\ref{JLC_iso}) with $K <0$ grow exponentially, $J(s) \propto \exp\left(\sqrt{-K} s\right)$. All the geodesics are unstable also on a manifold with non-constant sectional curvature, provided the latter is strictly negative \cite{DoCarmo:book}, and this explains why geodesic flows on negatively curved manifolds play such a relevant r\^{o}le in ergodic theory \cite{ergodictheory:book}; unfortunately, no similarly general result can be proved for manifolds of non constant positive curvature (or for manifolds whose curvature does not have a definite sign). The other case where Eq.\ (\ref{JLC}) becomes simple enough is that of two-dimensional manifolds, without any constraints on their curvature. In this case the Jacobi equation in a geodesic frame reads as
\beq
\frac{d^2 J}{ds^2} + K(s) J = 0\,,
\label{JLC_2}
\eeq 
where $J$ is the single component of the Jacobi field (remember that $J$ is orthogonal to $v$ so that the geodesic frame can be chosen such as $J$ has components only along one direction) and $K(s) = K(v,J)$ is the sectional curvature in the directions of $v$ and $J$. We see that Eq.\ (\ref{JLC_2}) is remarkably similar to the Jacobi equation for constant-curvature manifolds, but for that $K(s)$ can now vary along the geodesic. The solutions of Eq.\ (\ref{JLC_2}) can grow exponentially, implying instability of the geodesics, for two reasons: either $K < 0$ always, or the oscillations of $K$ induce, via a sort of parametric instability, an exponential growth of the envelope of $J(s)$. As first suggested in \cite{Pettini:pre1993}, where also some numerical examples were worked out, and further elaborated in \cite{pre1993,prl1995,pre1996}, this mechanism can be at work also in high-dimensional manifolds and can explain the origin of chaos in those systems whose curvatures are mainly positive. 

Starting from this idea, an effective scalar stability equation can be derived from the Jacobi equation (\ref{JLC}) under the assumption of {\em quasi-isotropy} of the manifold \cite{physrep2000}. Loosely speaking, quasi-isotropy means that the oscillations of the curvature along a geodesic are small as compared to the average curvature along the geodesic itself. Without entering technical details, when $N \to \infty$ this assumption and the use of the Eisenhart metric allow to approximate Eq.\ (\ref{JLC}) with the following scalar equation
\beq
\frac{d^2 \psi}{dt^2} + k(t) \psi = 0\,,
\label{JLC_eff}
\eeq 
where $t$ is time\footnote{The affine parametrization (\protect\ref{affine}) of the geodesics of the Eisenhart metric that correspond to dynamical trajectories allows to pass from arclength $s$ to time $t$.}, $\psi$ denotes any of the components of an effective Jacobi field, and $k(t)$ is an effective curvature (defined below). Equation (\ref{JLC_eff}) is formally very close to Eq.\ (\ref{JLC_2}), with a remarkable difference: it is a stochastic equation. The effective curvature $k(t)$ is a Gaussian stationary stochastic process whose probability distribution is defined by the equilibrium distribution of the Ricci curvature $K_R$ per degree of freedom, as calculated in the microcanonical ensemble. More precisely, the average of $k(t)$ is given by 
\beq
k_0 = \langle k \rangle = \frac{1}{N}\langle K_R \rangle_\mu= \frac{1}{N}\langle\triangle V \rangle_\mu  
\label{k_0}
\eeq
and its fluctuation is
\beq
\sigma^2_k = \langle k^2\rangle - \langle k \rangle^2 = \frac{1}{N}\left(\langle\triangle V^2 \rangle_\mu - \langle\triangle V \rangle^2_\mu\right) \,.
\label{sigma_k}
\eeq
In Eqs.\ (\ref{k_0}) and (\ref{sigma_k}), $\langle \cdot \rangle_\mu$ stands for the microcanonical average. Moreover, $k(t)$ is a $\delta$-correlated process:
\beq
\langle k(t_1)k(t_2) \rangle - \langle k(t_1) \rangle \langle k(t_2) \rangle = \tau \sigma^2_k \delta(t_2 - t_1)\,,
\eeq
where the correlation time scale $\tau$ can be self-consistently estimated as follows \cite{physrep2000}. First, two independent time scales $\tau_1$ and $\tau_2$ are identified as, respectively,
\beq
\tau_1 = \frac{\pi}{2\sqrt{k_0 + \sigma_k}}
\label{tau_1}
\eeq
and
\beq
\tau_2 = \frac{\sqrt{k_0}}{\sigma_k}\,.
\label{tau_2}
\eeq
The scale $\tau_1$ is the dominant time scale as long as $\sigma_k \ll k_0$; for $\sigma_k \to 0$ it is related to the period of the oscillations of the Jacobi field on a manifold with constant (positive) curvature $k_0$. The other scale $\tau_2$ dominates when the fluctuations grow, eventually until they become of the same order of magnitude of the average curvature. Then $\tau$ is obtained from $\tau_1$ and $\tau_2$ such that the shortest time scale dominates:
\beq
\tau^{-1} = \tau_1^{-1} + \tau_2^{-1}\,.   
\label{tau}
\eeq
Having completely defined the process $k(t)$, the average exponential growth rate of the solutions of Eq.\ (\ref{JLC_eff}) can be calculated\footnote{Actually one computes the exponential growth rate of the average of the solutions, which might be different from the average of the exponential growth rate of the solutions. See the discussion in Sec.\ \protect\ref{subsec_lyap}.} analytically \cite{physrep2000,pre1996}. Since, as recalled above, the Jacobi equation written for the Eisenhart metric coincides with the tangent dynamics equation (\ref{tang_dyn}), this directly yields a geometric estimate of the largest Lyapunov exponent $\lambda$ as
\begin{subequations}
\bea
\lambda\left(k_0,\sigma_k,\tau\right) &  = &  \frac{1}{2}\left(\Lambda - \frac{4k_0}{3\Lambda} \right) \\
\Lambda & = & \left[\sigma_k^2 \tau + \sqrt{\left(\frac{4k_0}{3}\right)^3 + \sigma_k^4\tau^2}\right]^{1/3}~.
\eea
\label{lambda}
\end{subequations}
Some remarks are in order as to Eq.\ (\ref{lambda}). First, it yields an estimate of the largest Lyapunov exponent in terms of the average and fluctuations of the effective curvature, once $\tau$ is defined as a function of $k_0$ and $\sigma_k$ via Eq.\ (\ref{tau}), without any free parameter. However, the above definition of $\tau$ is by no means a direct consequence of any theoretical result, but only a rough, physically based estimate. We have thus explicitly indicated the dependence of $\lambda$ on $\tau$ in Eq.\ (\ref{lambda}) because such an estimate might well be improved independently of the general geometric framework. In \protect\cite{pre1998} it was also considered as a free parameter to be fitted after comparison with numerical data. In this work we shall keep the estimate (\ref{tau}) for $\tau$ in order to have a parameter-free prediction for the Lyapunov exponent $\lambda$. Second, the ingredients $k_0$ and $\sigma_k$ entering Eq.\ (\ref{lambda}) can be calculated analytically whenever the microcanonical average and fluctuations of the Laplacian of the potential energy of the system under investigation can be calculated exactly; otherwise, they can be obtained via relatively short numerical simulations, in any case much shorter than the long integrations of the tangent dynamics equation (\ref{tang_dyn}) needed to directly calculate Lyapunov exponents, especially when the latter are small. Finally, when $\sigma_k \ll k_0$, that is when the quasi-isotropy assumption is more likely to be correct, Eq.\ (\ref{lambda}) yields
\beq
\lambda \propto \sigma_k^2\,,
\eeq
thus elucidating the deep relation between curvature fluctuations and dynamical instability.

In many cases Eq.\ (\ref{lambda}) yields good, if not very good, estimates of the largest Lyapunov exponent \cite{physrep2000}; in the case of the Fermi-Pasta-Ulam $\beta$-model (FPU-$\beta$), that is a chain of oscillators with quartic nonlinearity, it yields a prediction whose agreement with numerical data is excellent over eight orders of magnitude in energy and six orders of magnitude in $\lambda$ \cite{pre1996,physrep2000}. Being the theory based on the quasi-isotropy assumption, one expects it may fail when $\sigma_k > k_0$. Indeed, the quantitative agreement between the estimate (\ref{lambda}) and numerically calculated Lyapunov exponents  is typically worse when the fluctuations are large as compared to the average, although the order of magnitude of the estimate is often correct as well as the estimated dependence of $\lambda$ on the energy density \cite{pre1996,physrep2000,chaos2005}.

\subsection{Curvature and collective behaviour: the effective geometry of the landscape}
\label{sec_collective}

Apart from giving the basis of the Lyapunov exponent estimate (\ref{lambda}), the effective stability equation (\ref{JLC_eff}) suggests that the dynamics of the system can be seen as a geodesic flow on an effective two-dimensional manifold\footnote{According to Eq.\ (\ref{JLC_eff}), each degree of freedom behaves independently of the others so that everything goes as if it feels an effective low-dimensional manifold.} whose curvature fluctuates around an average value $k_0$, the amplitude of the curvature fluctuations being $\sigma_k$. This provides an intuitive geometric picture of the energy landscape: when the curvature fluctuations are small with respect to the average curvature, the landscape is very smooth and one expects the dynamics to be only weakly chaotic, while if the fluctuations of the curvature are of he same order of magnitude as the average the landscape is rough and chaos should be strong. Indeed, this is what happens in quite a few Hamiltonian systems with many degrees of freedom, the FPU-$\beta$ model being the paradigmatic example: a crossover between a weakly and a strongly chaotic regime, which has been referred to as the strong stochasticity threshold (SST) \cite{PettiniLandolfi:pra1990,PettiniCerrutiSola:pra1991},  occurs when the energy density $\varepsilon = E/N$ is increased and it does correspond to a crossover between a smooth effective manifold, where $\sigma_k \ll k_0$, and a rougher one, with $\sigma_k/k_0 \simeq {\cal O}(1)$ \cite{pre1993,chaos2005}.

Even more dramatic changes in the effective geometry of the landscape have been detected in systems undergoing thermodynamic phase transitions: for instance, in models with short-range interactions where the phase transition breaks a $O(n)$ symmetry, curvature fluctuations $\sigma_k$ exhibit a cusp-like behaviour at the transition \cite{prl1997,pre1998,prerap1999,CerrutiSolaClementiPettini:pre2000,physrep2000,chaos2005}. 
This suggests that a geometric counterpart of the phase transition may be found in the dependence of the effective geometry of the energy landscape on the energy density. A case of particular interest is that of the ferromagnetic mean-field $XY$ model, often referred to as the hamiltonian mean-field (HMF) model \cite{AntoniRuffo:pre1995}, that is a model of fully coupled planar classical spins with ferromagnetic interactions, where a mean-field ferromagnetic-paramagnetic transition occurs while increasing the energy density above a critical value $\varepsilon_c$. In this model the quantities $k_0$ and $\sigma_k$ describing the effective geometry of the landscape can be calculated analytically, as first done by Firpo in \cite{Firpo:pre1998} to estimate the Lyapunov exponent using Eq.\ (\ref{lambda}) (see also Ref.\ \cite{physrep2000} for a discussion of that result from the point of view of the present discussion). The calculation shows that the phase transition corresponds to a geometric transition from a manifold with positive curvature and nonvanishing curvature fluctuations to a flat manifold with zero curvature fluctuations. The effective average curvature $k_0$ decreases with energy and vanishes continuously at the transition, while the curvature fluctuations $\sigma_k$ increase with $\varepsilon$ as long as $\varepsilon < \varepsilon_c$ and discontinuously jump to zero at $\varepsilon_c$. Both $k_0$ and $\sigma_k$ vanish for $\varepsilon > \varepsilon_c$. As we shall see in the following, this model coincides with a particular limit of the SGR model we are going to study, so that it is of particular interest to the present work.

Before describing the SGR model and studying the effective geometry of its energy landscape, it is worth briefly mentioning some further developments which have been triggered by the geometric approach. First, constructing some abstract geometric models able to reproduce the cusp-like behaviour of the curvature fluctuations at a phase transition \cite{prl1997} and noticing that such behaviours occur also with different metric functions \cite{prerap1999}, it was conjectured that these geometric transitions were a consequence of deeper topology changes: this was explicity shown in the particular case of the mean-field $XY$ model \cite{jsp2003} and quite a few research activity followed (see \cite{Pettini:book,Kastner:rmp2008} and references therein). Second, the study of the effective energy landscape geometry of a toy model of a polymer has shown that the behaviour of curvature fluctuations allows to discriminate between polymers that have a proteinlike behaviour, collapsing to a well-defined native state, and polymers that collapse to a random state \cite{prl2006_2,pre2008,book2009}. Hence this tool seems to be useful also to characterize transitional phenomena in intrinsically finite systems like proteinlike polymers.

\section{The model}
\label{sec_model}

In this Section we shall describe the self-gravitating ring (SGR) model and its equilibrium statistical properties, in particular in the microcanonical ensemble. In Sec.\ \ref{subsec_sgr} we shall discuss the model; in Sec.\ \ref{subsec_micro} we shall discuss its equilibrium statistical behavior. 

\subsection{The self-gravitating ring}
\label{subsec_sgr}

The SGR model describes $N$ identical classical particles of mass $m$ constrained to move on a ring of radius $R$ and mutually interacting via a regularized gravitational potential. It was introduced in \cite{SotaEtAl:pre2001} and its Hamiltonian is
\beq
\label{hamiltonianasgr1}
{\cal H}_{\mathrm{SGR}}=\frac{1}{2mR^2}\sum _{i=1}^{N}P_{i}^2- \frac{1}{2}\sum _{i,j = 1}^{N}\frac{Gm^2}{\sqrt{2}R\sqrt{1-\cos \left(\vartheta _{i}-\vartheta _{j}\right)+\alpha }}\,,
\eeq
where $\vartheta_i \in (-\pi,\pi]$ is the angular coordinate of the $i$-th particle,  
\beq
P_i = mR^2\frac{d\vartheta_i}{dt}
\eeq
is its angular momentum, $R$ is the radius of the circle and $G$ is Newton's gravitational constant. The dimensionless constant $\alpha > 0$ provides the short-distance regularization: when $\vartheta_i - \vartheta_j \ll 1$ the interaction between the $i$-th and the $j$-th particle is effectively harmonic on a length scale of the order of 
\beq
d_\alpha = R\sqrt{2\alpha}~.
\eeq 
The SGR model can thus describe a self-gravitating system when $d_\alpha \ll R$, i.e., $\alpha \ll 1$. On the other hand, in the opposite limit $\alpha \to \infty$ the SGR model becomes equivalent\footnote{Provided the coupling constant is suitably rescaled with $\alpha$; otherwise, the allowed range of energy values of the model shrinks to zero.} to the ferromagnetic mean-field $XY$ model, or HMF model.

Following \cite{SotaEtAl:pre2001,jstat2010} we can define a characteristic time scale $\tau$ (the typical time a particle needs to go around the circle when all the others are uniformly distributed) as 
\beq
\tau = \sqrt{\frac{R^3}{GNm}}
\eeq
and dimensionless momenta as 
\beq
p_i = \frac{d\vartheta}{d\tau}~, ~~~i = 1,\ldots,N~.
\eeq
One can thus write a dimensionless Hamiltonian ${\mathcal{H}}$ as
\beq
\label{Ham}
{\mathcal{H}} = \frac{\tau^2}{mR^2} {\cal H}_{\mathrm{SGR}} = \frac{1}{2} \sum _{i=1}^{N}p_{i}^2 - \frac{1}{2N\sqrt{2}}\sum_{i,j=1}^{N}\frac{1}{\sqrt{1-\cos \left(\vartheta _{i}-\vartheta _{j}\right)+\alpha }}\,.
\eeq
The dimensionless Hamiltonian (\ref{Ham}) is extensive, due to the $\frac{1}{N}$ rescaling of the coupling between the particles. The Kac prescription for making extensive a long-range interaction \cite{CampaEtAl:physrep} is here obtained via a suitable choice of the adimensionalization procedure. From now on we will consider only the dimensionless Hamiltonian (\ref{Ham}).

\subsection{Microcanonical thermodynamics}
\label{subsec_micro}

The equilibrium thermodynamics of the SGR has been studied in \cite{TatekawaEtAl:pre2005} and later in \cite{Cesare:thesis} by a mean-field method, expected to be exact in the thermodynamic limit $N\to\infty$. The main features of the thermodynamics depend on the choice of the softening parameter $\alpha$ and can be summarized as follows. There is a phase transition between a homogeneous high-energy phase and an inhomogeneous low-energy phase, occuring at a critical energy density $\varepsilon_c$ weakly dependent on $\alpha$ (for instance, $\varepsilon_c \simeq -0.227$ for $\alpha = 1$ and $\varepsilon_c \simeq -0.32$ for $\alpha = 10^{-2}$). In the microcanonical ensemble the phase transition is discontinuous if $\alpha \lesssim 10^{-4}$ and continuous otherwise, i.e., there is a microcanonical tricritical point at $\alpha \simeq 10^{-4}$. There is inequivalence between the canonical and the microcanonical ensemble (and the system exhibits a region of microcanonical negative specific heat) for any $\alpha \lesssim 10^{-1}$; for larger values of $\alpha$, the transition is continuous in both ensembles, the region of microcanonical negative specific heat disappears, there is ensemble equivalence and the thermodynamics of the SGR is qualitatively similar to that of the mean-field $XY$ model.  

In the rest of the paper we shall consider three values of $\alpha$: $\alpha = 1$, where the thermodynamics is very close to that of the mean-field $XY$ model; $\alpha= 10^{-2}$, where there is a region of negative specific heat but the microcanonical phase transition is still continuous; and finally $\alpha= 3\times 10^{-5}$ where the microcanonical behaviour of the SGR is closer to that of a real self-gravitating system, with a large region of negative specific heat and a discontinuous phase transition. This choice of values of $\alpha$ allows to explore all the qualitatively different regimes exhibited by the system; moreover, for these values results on the thermodynamics, useful for validating our results, were obtained in \cite{Cesare:thesis}. 

\begin{figure}
\includegraphics[width=12cm,clip=true]{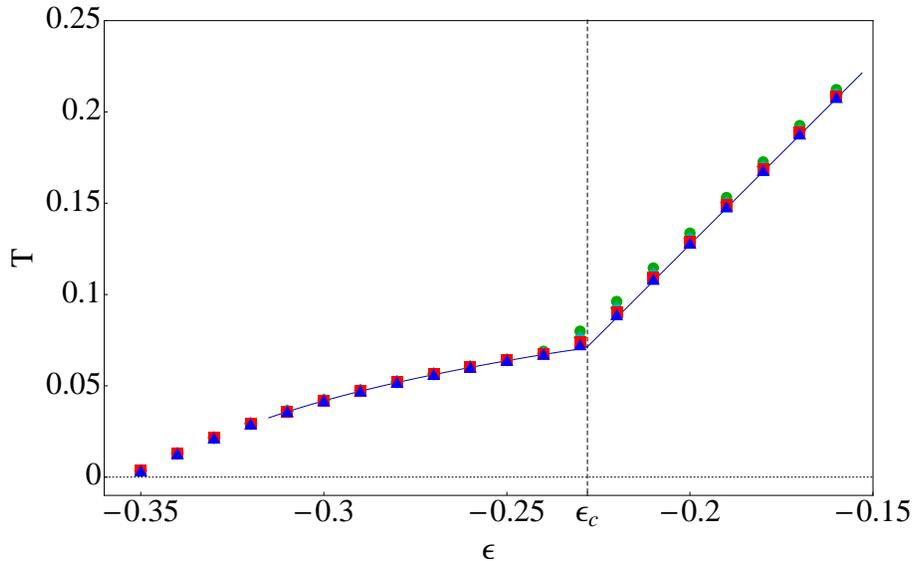}
\caption{(Color online) Microcanonical caloric curve $T(\varepsilon)$ of the SGR model for $\alpha = 1$. The blue solid line has been obtained in \protect\cite{Cesare:thesis} using the mean-field technique developed in \protect\cite{TatekawaEtAl:pre2005}; the symbols are MC results for $N=50$ (green circles), $N = 100$ (cyan rhombs), $N=200$ (red squares), and $N = 400$ (blue triangles). Statistical errors are smaller than the symbols' size. The dotted vertical line marks the critical energy density.}
\label{fig:caloric1}
\end{figure}

\begin{figure}
\includegraphics[width=12cm,clip=true]{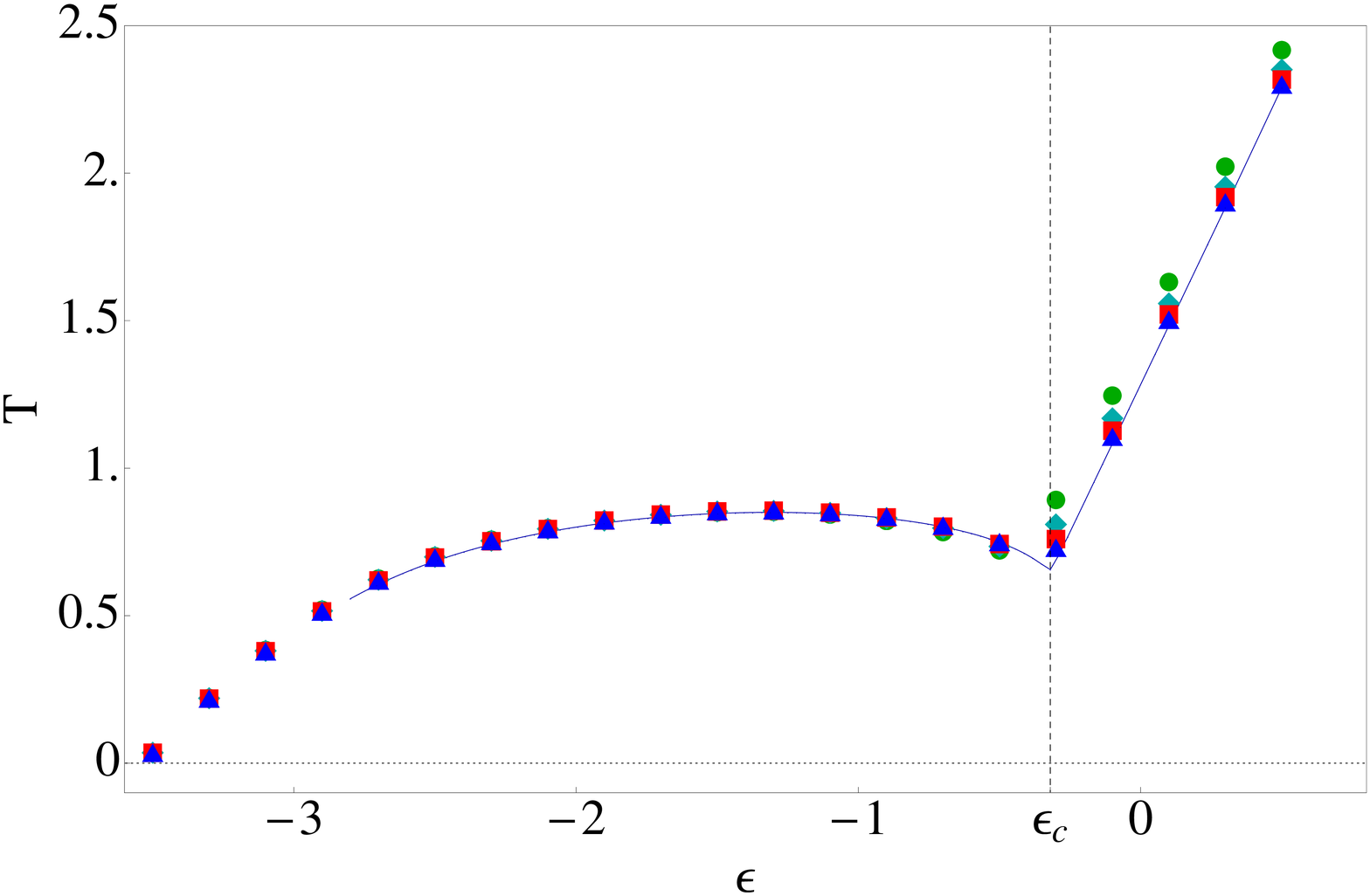}
\caption{(Color online) As in Fig.\ \protect\ref{fig:caloric1} for $\alpha = 10^{-2}$.}
\label{fig:caloric2}
\end{figure}

\begin{figure}
\includegraphics[width=12cm,clip=true]{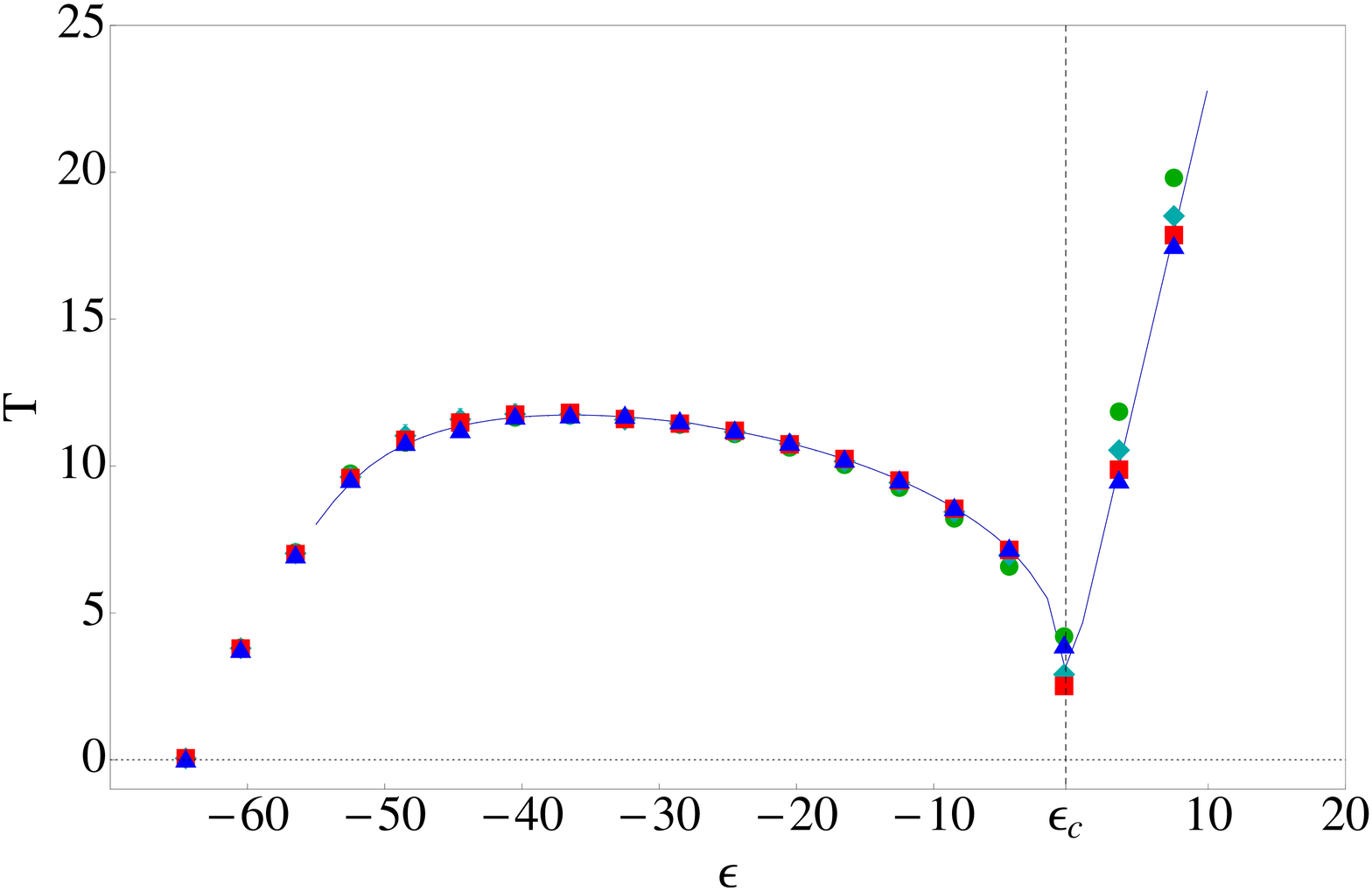}
\caption{(Color online) As in Fig.\ \protect\ref{fig:caloric1} for $\alpha = 3\times 10^{-5}$.}
\label{fig:caloric3}
\end{figure}

Since the mean-field technique developed in \cite{TatekawaEtAl:pre2005} does not allow to calculate fluctuations of thermodynamic observables, to compute $\sigma_k$ we had to resort to Monte Carlo (MC) simulations, performed using the Ray microcanonical algorithm \cite{Ray:pra1991}. In Figs.\ \ref{fig:caloric1},\ref{fig:caloric2} and \ref{fig:caloric3} we plot the caloric curves of the SGR as obtained with the mean-field method in \cite{Cesare:thesis} for the three chosen values of the softening parameter $\alpha$ and we compare them with the results obtained using the MC algorithm with $N$ ranging from 50 to 400, showing a very good agreement and the absence of appreciable $N$-dependence of the MC results in the inhomogeneous phase $\varepsilon < \varepsilon_c$. We note that MC results extend to smaller energies than those explored with the mean-field technique: the latter indeed does not easily converge for small energies.


\section{Results and discussion}
\label{sec_results}

Let us now present and discuss the results obtained for the average effective curvature $k_0$ and curvature fluctuations $\sigma_k$ of the energy landscape of the SGR. We shall first use these quantities to give a geometric characterization of the phase transition and then to estimate the largest Lyapunov exponent. Both quantities have been obtained by means of the MC method already mentioned in Sec.\ \ref{subsec_micro}; the average curvature $k_0$ has been calculated also using the mean-field technique of Ref.\ \cite{TatekawaEtAl:pre2005}, yielding results valid in the $N\to\infty$ limit. The number of particles considered in the MC simulations, as in the case of the computation of the caloric curves, ranged from $N = 50$ to $N = 400$. The SGR Hamiltonian (\ref{Ham}) is long-ranged and we did not use any cutoff or approximation on the forces as those typically used to simulate self-gravitating systems in astrophysical contexts (see e.g.\ \cite{BarnesHut:nature1986,BinneyTremaine:book}), so that the computational cost grows as $N^2$ and this sets a limit to the maximum number of particles that can be simulated in reasonable time. The computation of the caloric curves has shown that at least in the inhomogeneous phase the $N$-dependence of the MC results is practically negligible, so that we may expect that these numbers of particles are sufficient to yield a good estimate of the geometric properties of the landscape at large $N$. In all simulation runs we used ${\cal O}(10^6)$ MC steps per particle. 

In Sec.\ \ref{subsec_pt} we present the results for $k_0$ and $\sigma_k$ and discuss how they provide a geometric interpretation of the homogeneous-inhomogeneous phase transition. In Sec.\ \ref{subsubsec_2nd} we argue that these results may also support the existence of another transition in the SGR, at lower energies than $\varepsilon_c$, as suggested in \cite{prerap2009}. Finally, in Sec.\ \ref{subsec_lyap} we use $k_0$ and $\sigma_k$ to estimate the largest Lyapunov exponent of the SGR and we compare the theoretical estimates with direct numerical calculations.

\subsection{Geometry and phase transitions}
\label{subsec_pt}

\begin{figure}
\includegraphics[width=12cm,clip=true]{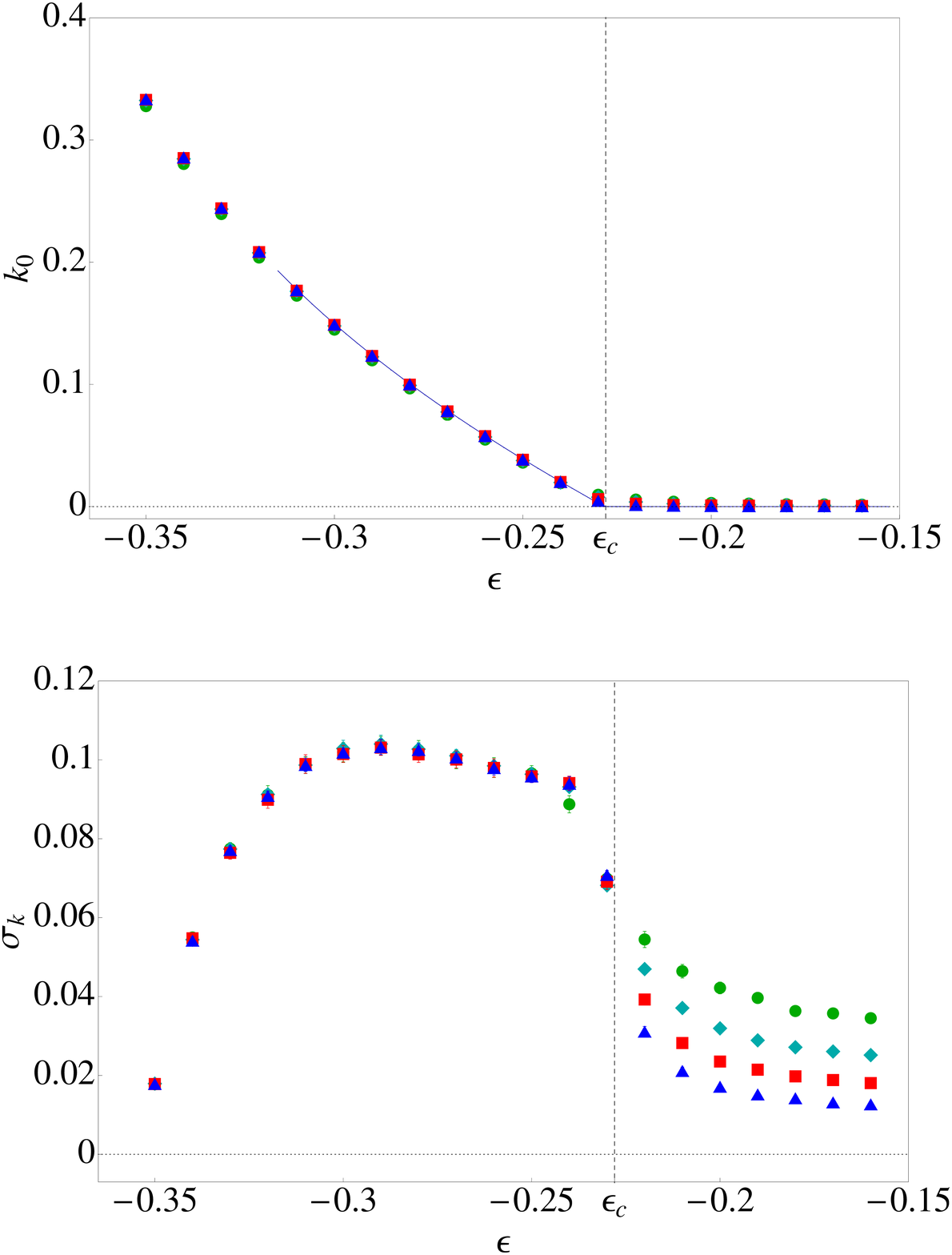}
\caption{(Color online) Average effective curvature $k_0$ (top panel) and effective curvature fluctuation $\sigma_k$ (bottom panel) of the energy landscape of the SGR as a function of the energy density $\varepsilon$ for $\alpha = 1$. Symbols refer to MC results for $N=50$ (green circles), $N = 100$ (cyan rhombs), $N=200$ (red squares), and $N = 400$ (blue triangles). Errorbars indicate statistical errors. The solid blue line in the upper panel is the mean-field result for $N\to\infty$. The dotted vertical line marks the critical energy density.}
\label{fig:geom1}
\end{figure}

\begin{figure}
\includegraphics[width=12cm,clip=true]{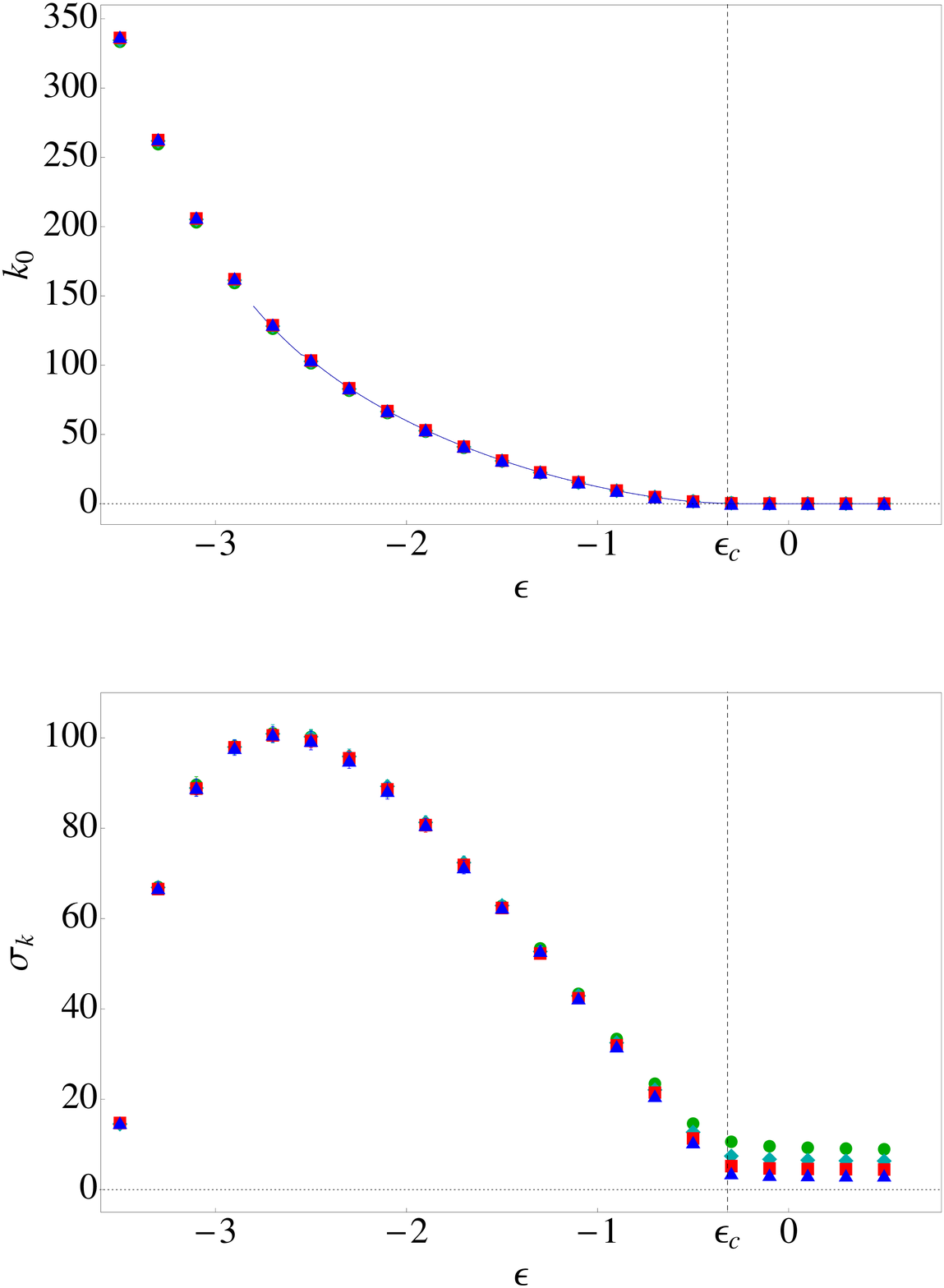}
\caption{(Color online) As in Fig.\ \protect\ref{fig:geom1} for $\alpha = 10^{-2}$.}
\label{fig:geom2}
\end{figure}

\begin{figure}
\includegraphics[width=12cm,clip=true]{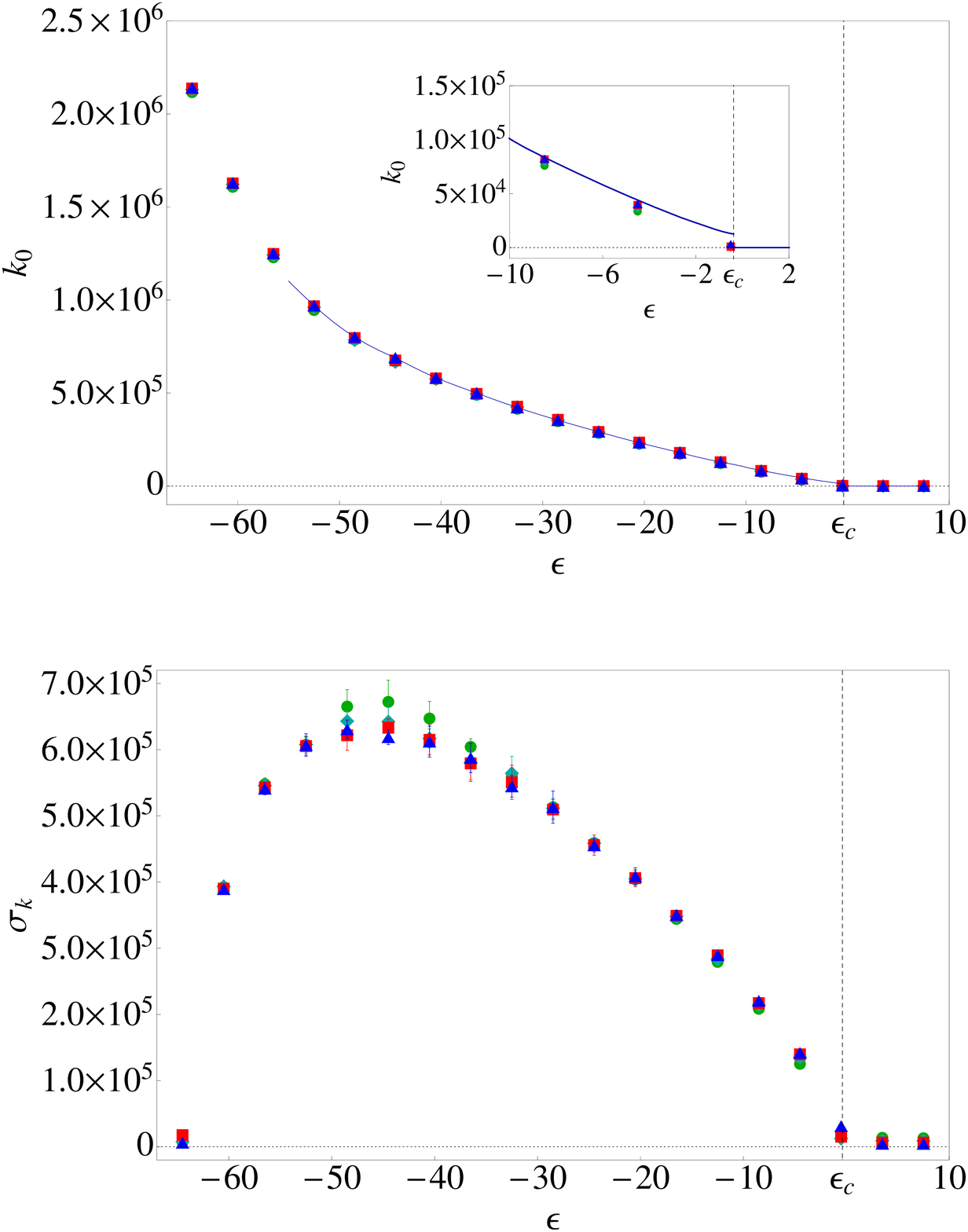}
\caption{(Color online) As in Fig.\ \protect\ref{fig:geom1} for $\alpha = 3\times 10^{-5}$. The inset in the top panel shows a magnified view close to the transition to appreciate the discontinuity in $k_0$ at $\varepsilon_c$ suggested by the mean-field results (blue solid line).}
\label{fig:geom3}
\end{figure}

The average effective curvature $k_0$ and fluctuations $\sigma_k$ for the SGR are given by inserting the potential energy of the Hamiltonian (\ref{Ham}) into the definitions (\ref{k_0}) and (\ref{sigma_k}); as already mentioned, the microcanonical averages have been obtained by MC simulations (and also using the mean-field method for $k_0$, at sufficiently high energies). The results for these quantities are reported in Figs.\ \ref{fig:geom1}, \ref{fig:geom2}, and \ref{fig:geom3}, for $\alpha = 1$, $\alpha = 10^{-2}$, and $\alpha = 3\times 10^{-5}$, respectively.


Starting from the case $\alpha = 1$ (Fig.\ \ref{fig:geom1}), we observe that the average curvature $k_0$ is positive, decreases with the energy density $\varepsilon$ and vanishes continuously at $\varepsilon_c$; it remains equal to zero in the whole homogeneous phase. More precisely, the mean-field method, valid as $N\to\infty$,  yields  $k_0 = 0$ $\forall \varepsilon \ge \varepsilon_c$; the MC data for $\varepsilon > \varepsilon_c$ are always positive but very small and exhibit a clear tendence to vanish as $N$ grows. For $\varepsilon < \varepsilon_c$ the MC data agree with the mean-field solution as long as the latter one can be reliably computed, and do not display relevant finite-size effects (in any case, these effects are smaller than in the homogeneous phase) as in the case of the caloric curves. The behaviour of the curvature fluctuations $\sigma_k$ is more complicated: $\sigma_k$ grows with $\varepsilon$ until a maximum is reached at $\varepsilon_{\text{max}}$ below $\varepsilon_c$, then it starts decreasing smoothly until, close to $\varepsilon_c$, a sharp decrease sets in and for $\varepsilon > \varepsilon_c$ the data show a strong $N$-dependence, compatible with $\sigma_k = 0$ for $N \to \infty$; indeed, in the homogeneous phase we have $\sigma_k(N) \propto N^{-\delta}$ with $\delta \simeq 0.5$. The data thus suggest that, as $N\to\infty$, $\sigma_k$ jumps discontinuously from a finite value to zero at $\varepsilon_c$. The qualitative behaviour of $k_0$ and $\sigma_k$ is thus very close to that observed in the mean-field $XY$ model (see Sec.\ \ref{subsec_pt}), the only relevant difference being that in the latter $\sigma_k$ reaches its maximum right at $\varepsilon_c$ where it also jumps to zero, while in the SGR the maximum occurs at $\varepsilon_{\text{max}}  < \varepsilon_c$. We have anyway considered also larger values of $\alpha$ (data not shown) and observed that $\varepsilon_{\text{max}}$ moves towards $\varepsilon_c$ as $\alpha$ grows, thus suggesting that at large values of $\alpha$, and ideally when $\alpha \to \infty$, the behaviour of $\sigma_k$ is exactly that of the mean-field $XY$ model, as expected.

Considering now the two smaller values of $\alpha$ (Figs.\ \ref{fig:geom2} and \ref{fig:geom3}), as far as $k_0$ is considered the qualitative behaviour is essentially the same as before, apart from a slight change in the shape of the curve $k_0(\varepsilon)$: $k_0 >0$ for $\varepsilon < \varepsilon_c$ and $k_0 = 0$ for $\varepsilon \ge \varepsilon_c$. However, while the transition between the two regimes seems continuous when $\alpha = 10^{-2}$ (Fig.\ \ref{fig:geom2}), our data are consistent with a discontinuous jump of $k_0$ at $\varepsilon_c$ for the smallest value of $\alpha$ we considered (Fig.\ \ref{fig:geom3}). The behaviour of the the curvature fluctuations $\sigma_k$ is instead very different from the $\alpha = 1$ case. Here, after reaching a maximum at an energy density $\varepsilon_{\text{max}}$ well below $\varepsilon_c$ ($\varepsilon_{\text{max}}$ of the same order of magnitude of, but not equal to, the energy density where the specific heat changes sign from positive to negative) $\sigma_k(\varepsilon)$ smoothly decreases to almost zero, and the residual jump at $\varepsilon_c$ is either absent, or, if present, is negligible as compared to the maximum value $\sigma(\varepsilon_{\text{max}})$. In the homogeneous phase, also for these values of $\alpha$ we have $\sigma_k(N) \propto N^{-\delta}$ with $\delta \simeq 0.5$. Apart from the already mentioned possible jump of $k_0$ at the transition for $\alpha = 3 \times 10^{-5}$, no other appreciable differences between $\alpha = 10^{-2}$ and $\alpha = 3 \times 10^{-5}$ show up in the graphs of $k_0(\varepsilon)$ and $\sigma_k(\varepsilon)$. There is a big change in the numerical values, that is however to be ascribed to the changes of the energy and length scales induced by the variation of $\alpha$. Hence a clear difference in the effective curvature of the energy landscape emerges between those values of $\alpha$ where the specific heat is always positive and there is equivalence between the canonical and microcanonical ensemble and those values where a negative specific heat region exists and the canonical and  microcanonical descriptions are no longer equivalent. Moreover, the fact that the microcanonical phase transition is discontinuous seems to be reflected in a discontinuity of $k_0$ at $\varepsilon_c$ when $\alpha = 3\times 10^{-5}$. 

The global geometric picture of the energy landscape that emerges from these results is very similar to that of the mean-field $XY$ model: the phase transition from an inhomogeneous to a homogeneous state corresponds to a transition from a positive-curvature manifold, with energy-dependent curvature fluctuations, to a flat manifold with no curvature fluctuations. The manifold corresponding to the homogeneous phase is easily identified with the whole configuration space, which is an $N$-torus $\mathbb{T}^N$, endowed with an effectively flat metric. This is true regardless of the value of $\alpha$. The differences emerge in the collapsed phase, and the data we have obtained suggest that these differences are related to how this torus is ``constructed'' starting from an energy landscape whose geometry is that of a sphere (positive curvature with vanishing curvature fluctuations) at small energy densities. Starting from the minimum energy density, when the latter increases the average curvature $k_0$ decreases while the curvature fluctuations increase until they reach a maximum at $\varepsilon_{\text{max}}$. The fact that $k_0$ decreases while $\sigma_k$ increases may be related to that larger and larger regions with smaller---or even zero or negative---local curvature are visited by the system and the effective manifold becomes somewhat intermediate between a sphere and a torus. This happens for any value of $\alpha$. Then, after the maximum the fluctuations starts to decrease; if $\alpha$ is sufficiently large, they suddenly jump to zero, as if the rest of the torus were constructed all of a sudden, by pinching all the remaining holes all together. Indeed, in Ref.\ \cite{jsp2003} a topological analysis of the configuration space of the mean-field $XY$ model showed that in the inhomogeneous phase a ``half-torus'' is progressively built, which becomes a full torus all of a sudden at the transition. For smaller values of $\alpha$ this sharp geometric transition is no longer present (or at least makes the curvature fluctuations change only by a small fraction of their maximum value): also the rest of the torus is progressively built, as its first half. The presence of a negative specific heat in thermodynamics appears thus to be related to a region where the ``second half'' of the torus, with more and more regions of zero or even negative curvatures, is progressively explored: this region is ``compressed'' in a sharper geometric change when the specific heat is always positive. It must be noted that if ascribing the decrease of the average curvature to the visiting of larger and larger regions of zero or negative curvature is correct, one would naturally expect that in the homogeneous phase the systems visits both positive and negative curvatures: if that occurs with equal probability, $k_0$ vanishes but $\sigma_k$ reaches a constant value, instead of vanishing. Indeed, this is what happens in $XY$ models with short range-interactions \cite{pre1996,CerrutiSolaClementiPettini:pre2000}, whose configuration space is also $\mathbb{T}^N$. In our case we have $k_0 = 0$ but also $\sigma_k = 0$, so that it seems that in the homogeneous phase the system visits only regions with zero curvatures, or that the metric becomes flat on the whole manifold. The fact that the effective energy landscape becomes flat in the homogeneous phase is thus an effect of the long-range nature of the interactions. 

\begin{figure}
\includegraphics[width=12cm,clip=true]{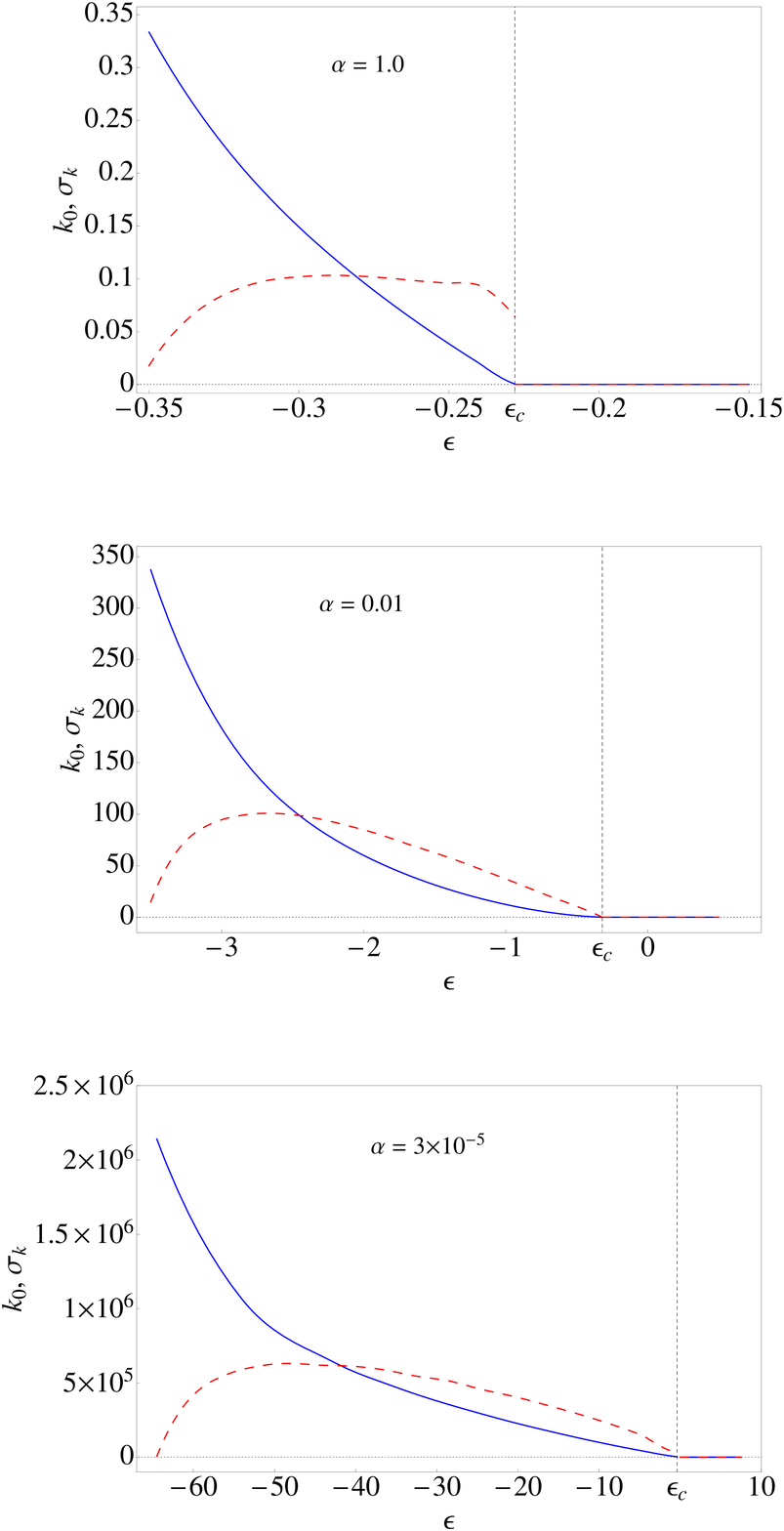}
\caption{(Color online) Estimate of the effective geometry of the energy landscape of the SGR extrapolated to $N\to\infty$ as a function of the energy density $\varepsilon$ for $\alpha = 1$ (top panel), $\alpha = 10^{-2}$ (middle panel), and $\alpha = 3 \times 10^{-5}$ (bottom panel). The blue solid line is the average effective curvature $k_0$ and the red dashed line is the effective curvature fluctuation $\sigma_k$. The curves have been obtained by mean-field results where available (that is, for $k_0$ for all the values of $\varepsilon$ but the smaller ones), by interpolating the data with $N=400$ in the inhomogeneous phase and by setting $\sigma_k$ to zero in the homogeneous phase. The dotted vertical line marks the critical energy density.}
\label{fig:geom4}
\end{figure}

As a final remark, it is important to note that for all the values of $\alpha$ there exists an energy region in the inhomogeneous phase where $\sigma_k > k_0$. The latter region extends from an energy density slightly smaller than $\varepsilon_{\text{max}}$ to $\varepsilon_c$, as shown in Fig.\ \ref{fig:geom4}, where the extrapolation of the results on $k_0$ and $\sigma_k$ to $N\to\infty$ is reported. The presence of such a region has two main consequences. First, it means that the effective geometry of the landscape is particularly complicated in that region, since fluctuations of the curvature larger than the average curvature itself show up on the macroscopic scale. Second, in this region the hypothesis of quasi-isotropy of the manifold at the basis of the geometric estimate (\ref{lambda}) of the Lyapunov exponent $\lambda$ no longer holds, and we may expect this reflects in a disagreement between estimated and directly calculated values of $\lambda$. As we shall see in Sec.\ \ref{subsec_lyap}, this is precisely what happens. But before considering Lyapunov exponents, let us concentrate on another aspect of the effective geometry of the landscape.

\subsubsection{Another transition?}
\label{subsubsec_2nd}

In Ref.\ \cite{prerap2009} the stationary points of the energy landscape of the SGR have been studied, showing that the homogeneous-inhomogeneous phase transition is related to the fact that a class of stationary points becomes asymptotically flat, that is the determinant of the Hessian of the potential energy vanishes for those stationary points when $N\to\infty$. The asymptotic flatness had been proposed as a necessity criterion for selecting stationary points able to induce a phase transition in \cite{KastnerSchnetz:prl2008,KSS:jstat2008} and has been later on referred to as the KSS criterion. Being based on stationary points of the potential energy landscape, the KSS criterion singles out a value of the potential energy density $v = V/N$ where a phase transition can occur. In Ref.\ \cite{prerap2009} it was shown that not only $v = 0$, that is the maximum of the potential energy density, corresponding to the homogeneous phase, satisfies the KSS criterion, but also another value, which depends on the softening parameter $\alpha$:
\beq
v_c = -\frac{4 + \alpha[6 + \alpha(5+2\alpha)]}{\sqrt{2\alpha}\left[(2+\alpha)^{3/2} + \alpha^{3/2}\right]}~.
\label{vc}
\eeq 
Hence another phase transition might be present in the SGR, at a critical energy such that the average value of the potential energy density equals $v_c$. 
It was also suggested that this transition, if present, could separate a low-energy region where the probability that a particle goes around the circle drops to zero (so that the system has a core-only density profile) from a higher-energy region where the probability of exploring the whole circle becomes nonzero, although the phase is still clustered (so that the system has a core-halo density profile). Such a change in the distribution may have very small effects on the thermodynamics. In a recent work \cite{RochaFilhoEtAl:pre2011} no signs of such a transition have been detected in the thermodynamic observables of the SGR, but indeed a change in the density profile of the same kind as anticipated above has been observed. A similar behaviour has been observed in a toy model that well approximates the SGR when $\alpha$ is small (see Ref.\ \cite{jstat2010}, where it was also shown that the same happens in the Thirring model \cite{Thirring:zphys1970}).

\begin{figure}
\includegraphics[width=9cm,angle=-90,clip=true]{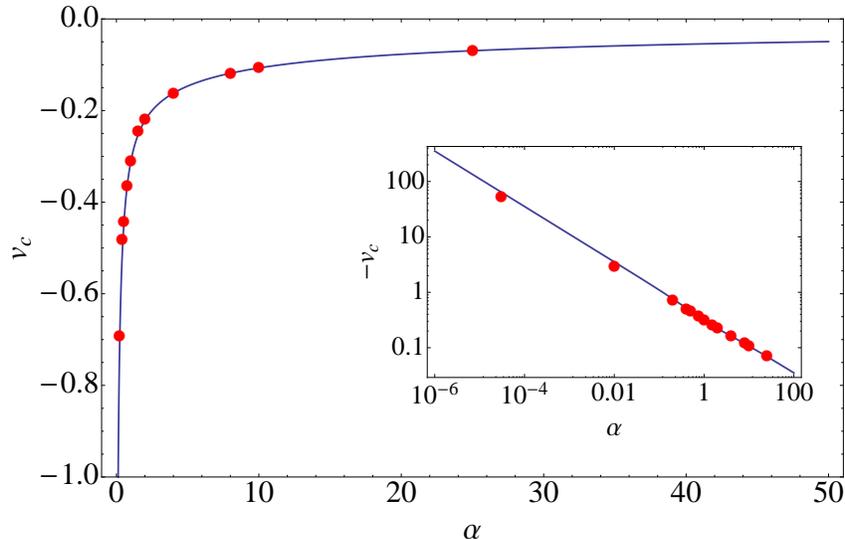}
\caption{(Color online) Comparison between the potential energy density at the maximum of the curvature fluctuations $v_{\text{max}}$ defined in Eq.\ (\protect\ref{vmax}) (red symbols) and  the potential energy density $v_c$ predicted by the KSS criterion as given by Eq.\ (\protect\ref{vc}) (blue solid line) as a function of the softening parameter $\alpha$. Inset: the same in a log-log scale, to show a wider range of values of $\alpha$.}
\label{fig:2nd}
\end{figure}

We are now going to show that at $v_c$ something happens also in the effective geometry of the energy landscape, and precisely a maximum in the curvature fluctuations. We have already noted that the curvature fluctuations $\sigma_k$ show a maximum at an energy density $\varepsilon_{\text{max}}$ which is always smaller than $\varepsilon_c$ but tends to the latter value when $\alpha \to \infty$. On the other hand, previous studies have shown that a maximum in the curvature fluctuations appears to be often (if not always) associated to big changes in the collective behaviour of the system: typically to a thermodynamic phase transition \cite{physrep2000} (in this case the curvature fluctuations appear to be singular at the maximum, e.g.\ showing a a cusp) or to a big configurational change which is not a thermodynamic transition due to the finiteness of the system, as in the toy model of a proteinlike polymer studied in \cite{prl2006_2,pre2008}. In the latter case the maximum of the curvature fluctuations is smooth. The maximum of the curvature fluctuations of the SGR does not show a tendency to become singular as $N$ grows, so that in a sense it seems more similar to that of the polymer. It remains to show that the maximum of the curvature fluctuations occurs at $v_c$; in Fig.\ \ref{fig:2nd} we plot the microcanonical average of the potential energy density at the energy density $\varepsilon_{\text{max}}$ of the maximum of the curvature fluctuations, 
\beq
v_{\text{max}} = \frac{1}{N}\langle V \rangle_\mu \left( \varepsilon_{\text{max}}\right)\,,
\label{vmax}
\eeq
as a function of the softening parameter $\alpha$, together with the predicted value $v_c$ given by Eq.\ (\ref{vc}). The agreement is quite good. This result suggests that a deep change in the properties of the system, which as observed in \cite{RochaFilhoEtAl:pre2011} is probably not a usual thermodynamic transition but a change from a core-only to a core-halo density profile, actually occurs at $v_c$. We also observe that this confirms once again the sensitivity of the effective geometry of the energy landscape to changes in the collective properties which may not affect usual thermodynamic observables like temperature, internal energy or specific heat, as it happens for proteinlike polymers \cite{prl2006_2,pre2008}.

\subsection{Geometry and chaos}
\label{subsec_lyap}

We can now use the results on the average effective curvature $k_0$ and curvature fluctuations $\sigma_k$ presented in Sec.\ \ref{subsec_pt} to estimate the largest Lyapunov exponent of the SGR using Eq.\ (\ref{lambda}). 

\begin{figure}
\includegraphics[width=12cm,clip=true]{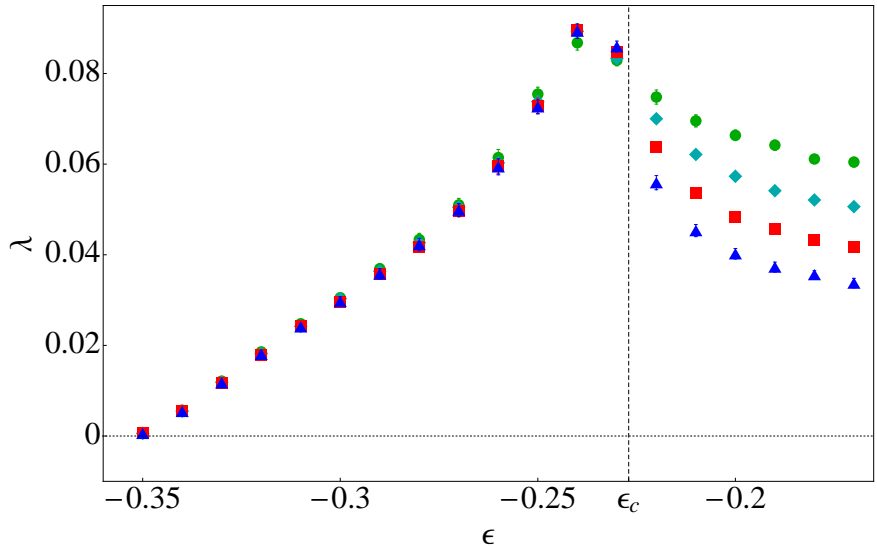}
\caption{(Color online) Geometric estimate of the largest Lyapunov exponent $\lambda$ of the SGR as a function of the energy density $\varepsilon$ for $\alpha = 1$, obtained inserting MC results for $k_0$ and $\sigma_k$ into Eq.\ (\protect\ref{lambda}) with $N=50$ (green circles), $N = 100$ (cyan rhombs), $N=200$ (red squares), and $N = 400$ (blue triangles). Errorbars indicate statistical errors. The dotted vertical line marks the critical energy density.}
\label{fig:lyapgeom}
\end{figure}

First of all, we note that our results, obtained with $N \leq 400$, are consistent with the expectation that the dynamics is regular in the thermodynamic limit in the homogeneous phase, i.e., $\lambda \to 0$ when $N\to\infty$ and $\varepsilon \ge \varepsilon_c$, since both $k_0$ and $\lambda$ vanish in the thermodynamic limit. More precisely, using Eq. (\ref{lambda}), when $\varepsilon \ge \varepsilon_c$ we obtained $\lambda(N) \propto N^{-\gamma}$, with $\gamma \simeq 0.28$ for $\alpha = 1$,  $\gamma \simeq 0.31$ for $\alpha = 10^{-2}$, and $\gamma \simeq 0.38$ for $\alpha = 3\times 10^{-5}$. These $N$-dependencies compare reasonably well with theoretical estimate made in the case of the mean-field $XY$ model \cite{Firpo:pre1998}, that is $\gamma = 1/3$. At variance with the homogeneous phase, the $N$-dependence of the geometric estimate of $\lambda$ in the inhomogeneous phase $\varepsilon < \varepsilon_c$ is weak, as a consequence of the weak $N$-dependence of $k_0$ and $\sigma_k$. We can thus assume that the estimate of $\lambda$ obtained inserting in Eq.\ (\ref{lambda}) our MC results for $k_0$ and $\sigma_k$ with $N=400$ is a reasonable estimate also for $N\to\infty$, in the collapsed phase. In Fig.\ \ref{fig:lyapgeom} we show the geometric estimate of $\lambda$ obtained from Eq.\ (\ref{lambda}) for $\alpha = 1$ and $N$ ranging from 50 to 400. 

\begin{figure}
\includegraphics[width=9cm,angle=-90,clip=true]{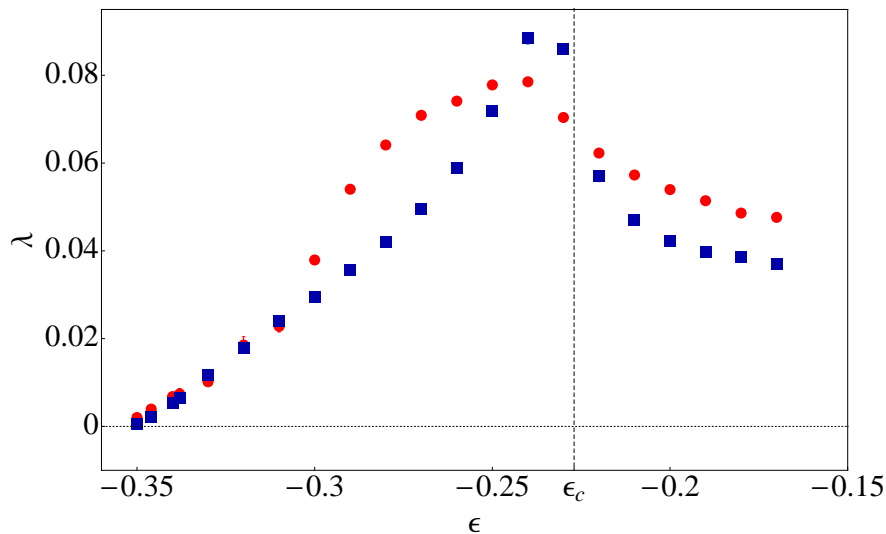}
\caption{(Color online) Comparison between the geometric estimate (\protect\ref{lambda}) of the largest Lyapunov exponent $\lambda$ (blue squares) and the direct measurement of $\lambda$ obtained by molecular dynamics simulations (red circles), as a function of the energy density $\varepsilon$ for $\alpha = 1$ and $N = 400$. Errorbars indicate statistical errors. The dotted vertical line marks the critical energy density.}
\label{fig:lyap1}
\end{figure}
\begin{figure}
\includegraphics[width=9cm,angle=-90,clip=true]{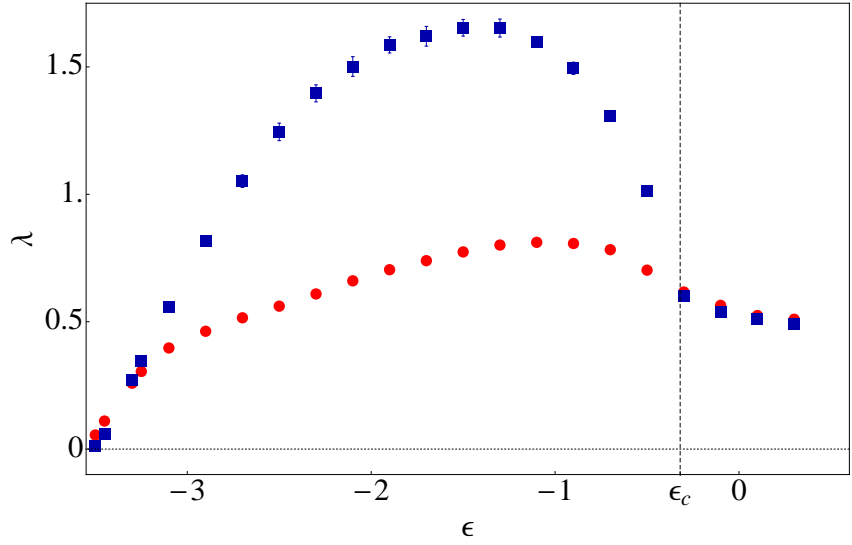}
\caption{(Color online) As in Fig.\ \protect\ref{fig:lyap1}, with $\alpha = 10^{-2}$.}
\label{fig:lyap2}
\end{figure}
\begin{figure}
\includegraphics[width=9cm,angle=-90,clip=true]{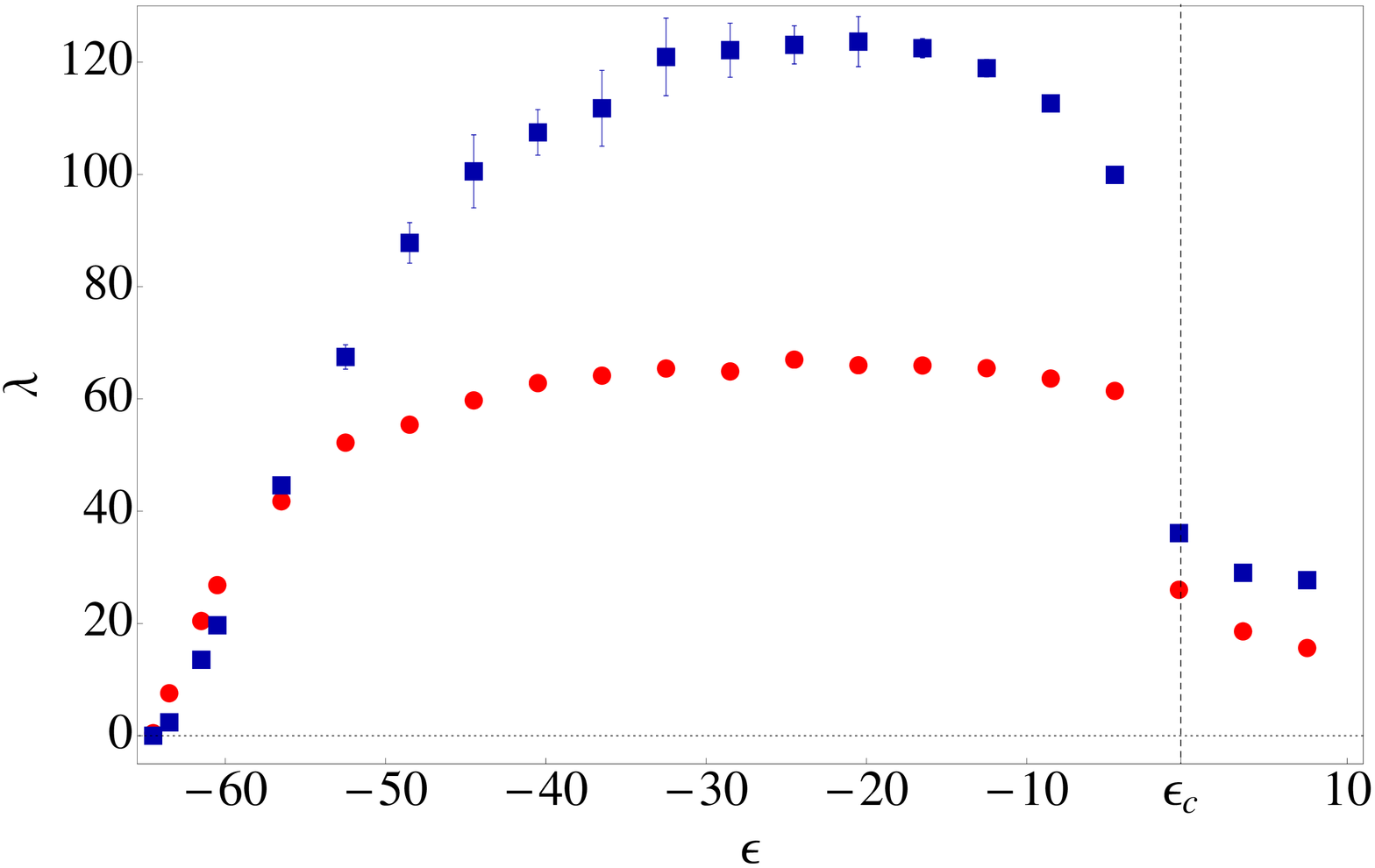}
\caption{(Color online) As in Fig.\ \protect\ref{fig:lyap1}, with $\alpha = 3 \times 10^{-5}$ and $N=100$.}
\label{fig:lyap3}
\end{figure}

In Figs.\ \ref{fig:lyap1}, \ref{fig:lyap2} and \ref{fig:lyap3} we compare the geometric estimates of $\lambda$ for $\alpha = 1$, $\alpha = 10^{-2}$ and $\alpha = 3\times 10^{-5}$, respectively, with direct measurements of $\lambda$ obtained by means of molecular dynamics (MD) simulations. For the sake of clarity, only the results for the largest number of particles considered are reported for each value of $\alpha$, since the geometric theory is expected to perform better for larger systems\footnote{By the way, in the present case the quantitative agreement between geometric estimates and numerical values of $\lambda$ turned out to be comparable for all the $N$'s considered, and the largest $N$ did not necessarily yield the best result.}. The MD simulations were performed by numerically integrating both the equations of motion derived from the SGR Hamiltonian (\ref{Ham}) and the tangent dynamics equation (\ref{tang_dyn}), i.e., the equation of motion in the tangent space obtained by linearizing the equations of motion along the dynamical trajectory. Numerical integration was performed using the McLachlan-Atela symplectic integrator \cite{McLachlanAtela:nonlinearity1992} and relative fluctuations of the energy remained smaller than $10^{-5}$ in the worst case. The computational cost of the MD simulations of the SGR grows as $N^2$ as the MC one, but is even higher because to compute Lyapunov exponents one has to simulate both the dynamics in the configuration space and that in the tangent space. Moreover to extract reliable values of $\lambda$ with the standard method \cite{BenettinEtAl:Meccanica1980} very long simulations are needed. We were thus able to obtain reliable direct numerical estimates of $\lambda$ for systems up to $N=400$ with $\alpha = 1$ and $\alpha=10^{-2}$, but only up to $N=100$ with $\alpha = 3\times 10^{-5}$. Although only the results for the largest $N$'s are shown, also the direct MD estimates of $\lambda$ are consistent with $\lambda \to 0$ when $N\to \infty$ and $\varepsilon \ge \varepsilon_c$, at least for the two larger values of $\alpha$ considered; for $\alpha = 1$ we found $\lambda (N) \propto N^{-0.2}$ and $\lambda (N) \propto N^{-0.25}$ for $\alpha = 10^{-2}$. For $\alpha = 3\times 10^{-5}$ we were unable to extract the $N$-dependence of $\lambda$. We also note that a residual $N$-dependence of the MD values of $\lambda$ shows up also for $\varepsilon \lesssim \varepsilon_c$, at variance with the geometric estimates. The latter $N$-dependence is larger at smaller $\alpha$'s. For small $\varepsilon$, no appreciable $N$-dependence was observed.

Several observations are in order as Figs.\ \ref{fig:lyap1}, \ref{fig:lyap2} and \ref{fig:lyap3} are concerned. First, the order of magnitude of the geometric estimate (\ref{lambda}) always compares very well with the MD data, for any $\alpha$ and any $\varepsilon$, the worst difference between geometric estimates and MD data being a factor of 1.5 for $\alpha = 1$ and a factor of 2 for $\alpha = 10^{-2}$ and $\alpha = 3\times 10^{-5}$. This is far from being trivial, since the absolute values of $\lambda$ vary by more than a factor of 100 passing from $\alpha = 1$ to $\alpha = 3\times 10^{-5}$. The same can be said of the overall behaviour of the curves $\lambda(\varepsilon)$, which is reasonably reproduced by the geometric estimates, although the agreement is worse as $\alpha$ gets smaller. Second, the agreement between geometric estimates and MD data becomes quantitatively very good at small $\varepsilon$ for all the $\alpha$'s considered, and also in the whole homogeneous phase $\varepsilon \ge \varepsilon_c$ if we assume that the extrapolation $\lambda \to 0$ for large systems is correct. These two situations are precisely those where the theory leading to Eq.\ (\ref{lambda}) is expected to be correct. In the homogeneous phase the correctness of the geometric theory as $N\to\infty$ is somewhat trivial: geodesics of a flat manifold are not chaotic. However, also at finite $N$ the geometric estimate yields reasonable results in the homogeneous phase. At small $\varepsilon$ one has $\sigma_k \ll k_0$ so that the quasi-isotropy assumption is consistent and the theory performs very well as expected. Third, in an intermediate energy region there is a quantitative disagreement between theory and direct numerical measurements. As already pointed out in Sec.\ \ref{subsec_pt}, in this region $\sigma_k \gtrsim k_0$ so that the quasi-isotropy assumption is questionable and a disagreement is not surprising at all, and has been observed in all the cases where curvature fluctuations grow too much with respect to the average curvature (see e.g.\ \cite{physrep2000}). The more so, because $\sigma_k > k_0$ means that negative curvatures appear with non-negligible probability, and the combined effect of the parametric instability due to fluctuations and the defocusing effect due to negative curvature is often very difficult to understand, resulting sometimes in a reinforcement of the overall chaoticity and sometimes in an inhibition of chaos (see e.g.\ \cite{Pettini:book}). In some cases as that of the 1-$d$ $XY$ model \cite{pre1996} the geometric estimate of $\lambda$ can be corrected to account for the negative curvatures' contribution, yielding a very good agreement between theory and numerics also when $\sigma_k > k_0$. We did not try to implement a correction as such in the present case, nor did we try to use the correlation timescale $\tau$, estimated in Eq.\ (\ref{tau}), as a fitting parameter as done in \cite{pre1998}.

Finally, we note that the quantitative disagreement between theoretical estimates and numerical measurements of $\lambda$ when $\sigma \gtrsim k_0$ may also have a different origin. As correctly pointed out by Vallejos and Anteneodo in \cite{VallejosAnteneodo:pre2012} and also by Politi \cite{Politi:privcomm}, Eq.\ (\ref{lambda}) does not directly estimate the Lyapunov exponent $\lambda$ but rather a generalized Lyapunov exponent commonly referred to as $\lambda_2$. This is due to that, for ease of calculation, in deriving Eq.\ (\ref{lambda}) one first averages the trajectories of the stochastic oscillator over the realizations of the stochastic process and then computes the exponential growth rate of the averaged solution. To properly compute $\lambda$ one should instead first compute the exponential growth rate of the solutions and then take the average over the realizations (this is formally very similar to the difference between quenched and annealed averages in the statistical physics of disordered systems). When the amplitude of the fluctuations is small $\lambda_2$ is typically a very good estimate of $\lambda$, but this is not always the case when the amplitude of the fluctuations get larger. However, the pursuit of this idea is left for future work.

\section{Summary and concluding remarks}
\label{sec_conclusions}

We have studied the global geometry of the energy landscape of the SGR by endowing the configuration space with the Eisenhart metric, such that the dynamical trajectories are identified with geodesics. The average curvature and curvature fluctuations of the energy landscape have been computed by means of MC simulations and, when possible, of a mean-field method, showing that these global geometric quantities provide a geometric characterization of the collapse phase transition. In particular, the average curvature $k_0$ as a function of the energy density $\varepsilon$ behaves as an order parameter, vanishing for $\varepsilon \ge \varepsilon_c$. Curvature fluctuations $\sigma_k$ provide further insight into the nature of the collapsed phase and on its dependence on the softening parameter $\alpha$. The behaviour of $\sigma_k(\varepsilon)$ also shows a maximum in correspondence with the energy of a possible further transition, occurring at lower energies, whose existence had been previously conjectured on the basis of a local analysis of the energy landscape \cite{prerap2009}. Such a transition may correspond to a change from a a core-only to a core-halo density profile, although it does not appear to have appreciable consequences on the usual thermodynamic quantities \cite{RochaFilhoEtAl:pre2011}. This confirms once more the sensitivity of the geometric observables to changes in the collective properties which may not affect usual thermodynamic observables like temperature, internal energy or specific heat, as it happens for instance in the case of the proteinlike polymers considered in \cite{prl2006_2,pre2008}. 

The effective curvature and its fluctuations allowed us to estimate the largest Lyapunov exponent $\lambda$ of the SGR via Eq.\ (\ref{lambda}). The geometric estimate always gives the correct order of magnitude of $\lambda$ and is also quantitatively correct at small $\varepsilon$ and, in the limit $N\to\infty$, in the whole homogeneous phase $\varepsilon \ge \varepsilon_c$. For intermediate energies the accuracy of the estimate is worse. Though not surprising at all, since for these values of the energy density the assumptions leading to Eq.\ (\ref{lambda}) may fail, this result nonetheless calls for a deeper understanding which may also lead to an improvement of the accuracy of the geometric approach to chaos, as discussed in Sec.\ \ref{subsec_lyap}. Even if, in our opinion, the main value of the geometric approach is in its explanation of the deep origin of chaos in Hamiltonian many-particle systems and not in its effectiveness as a computational tool for Lyapunov exponents, it is also true that any theory is valid inasmuch as it can predict the outcomes of (numerical, in the present case) experiments. Hence the possibility of modifying Eq.\ (\ref{lambda}) to incorporate a better estimate of $\lambda$ is a very interesting suggestion for future work.

To conclude, all the results on the effective geometry of the energy landscape of the SGR show how this global approach to the properties of the energy landscape may be useful and complementary to the local analysis of stationary points, as that carried out in \cite{prerap2009} for the same model.

\acknowledgments
We warmly thank C.\ Nardini for many discussions, for providing us with some useful codes and results and for his interest in our work. We also thank P.\ Bernab\`o for extensive testing of an earlier version of the MC code.

\bibliography{/Users/casetti/Work/Scripta/papers/bib/mybiblio,/Users/casetti/Work/Scripta/papers/bib/statmech,/Users/casetti/Work/Scripta/papers/bib/astro}

\end{document}